\documentclass[authoryear,12pt,5p,times]{elsarticle}

\usepackage{bm}
\usepackage{xspace}

\newcommand{\PlP}{{\sc PlanetPack}\xspace}

\newcommand{\expect}{\mathbb{E}}

\newcommand{\var}{\mathop{\text{Var}}\nolimits}
\newcommand{\cov}{\mathop{\text{Cov}}\nolimits}

\usepackage{amssymb}
\usepackage{amsmath}

\journal{Astronomy and Computing}

\begin{document}
\sloppy

\begin{frontmatter}



\title{\textbf{PlanetPack3}: a radial-velocity and transit analysis tool for exoplanets}


\author{Roman V. Baluev}

\address{Central Astronomical Observatory at Pulkovo of Russian Academy of Sciences,
Pulkovskoje sh. 65/1, Saint Petersburg 196140, Russia}
\address{Saint Petersburg State University, Faculty of Mathematics \& Mechanics, Universitetskij
pr. 28, Petrodvorets, Saint Petersburg 198504, Russia}
 \ead{r.baluev@spbu.ru}

\begin{abstract}
PlanetPack, initially released in 2013, is a command-line software aimed to facilitate
exoplanets detection, characterization, and basic dynamical $N$-body simulations. This
paper presents the third major release of PlanetPack that incorporates multiple
improvements in comparison to the legacy versions.

The major ones include: (i) modelling noise by Gaussian processes that in addition to the
classic white noise may optionally include multiple components of the red noise, modulated
noise, quasiperiodic noise (to be added soon in minor subversions of the 3.x series); (ii)
an improved pipeline for TTV analysis of photometric data that includes quadratic
limb-darkening model and automatic red-noise detection; (iii) self-consistent joint fitting
of photometric + radial velocity data with full access to all the functionality inherited
from the legacy PlanetPack; (iv) modelling of the Rossiter-McLaughlin effect for arbitrary
eclipser/star radii ratio, and optionally including corrections that take into account
average characteristics of a multiline stellar spectrum; (v) speed improvements through
multithreading and CPU-optimized BLAS libraries.

PlanetPack was written in pure C++ (standard of 2011), and is expected to be run on a wide
range of platforms.
\end{abstract}


\end{frontmatter}


\section{Introduction}
\label{sec_intro}
\PlP was initially released in 2013 \citep{Baluev13c}, targeting tasks of exoplanets
detection and characterization, based on Doppler radial-velocity (RV) data, and of
exoplanetary dynamics. Along its 1.x release series, this software offered the following
functionality: (i) RV curve fitting with a fittable RV jitter; (ii) fitting the RV data
with red noise (auto-correlated errors); (iii) multi-Keplerian as well as Newtonian
$N$-body RV fits; (iv) advanced maximum-likelihood periodograms; (v) calculation of
parametric confidence regions; (vi) constrained fitting; (vii) analytical statistical tests
and numerical Monte Carlo simulations; (viii) basic tasks of long-term planetary $N$-body
simulation.

The first major release of \PlP and the subsequent 1.x series were capable to deal with
only RV data. The second major release introduced a dedicated pipeline that allowed to
perform a homogeneous fit of multiple transit lightcurves of the same exoplanet
\citep{Baluev15a}, but this pipeline was largerly experimental that time. Moreover, it
wasn't yet possible in the 2.x series to analyse photometric and Doppler data together, so
this transit fitting pipeline still looked like a standalone foreign module inside the RV
analysis software.

However, the importance of the joint analysis of the RV+transit data is growing. Several
obvious reasons for that are listed below.
\begin{enumerate}
\item Transit and Doppler methods are highly complementary, so they provide much more
complete description of the planetary system when used together. This includes the
determination of the true planetary mass (resolving the inclination $i$ in the famous
$m\sin i$ issue), information about the 3D structure of the system if the star is transited
by multiple planets, estimation of planet density, high-accuracy determination of orbital
eccentricities.\footnote{The timing of a planetary transit yields a high-accuracy reference
phase on the Doppler curve. At this phase, the radial velocity deviation is proportional to
$\sim e\cos\omega$. Hence, this combination becomes constrained with a much better accuracy
if just a single transit is available in addition to radial velocities.}
\item Transit timing variations (TTVs) can provide independent hints of $N$-body
interactions in a planetary system. This method is even capable to detect previously
unknown planets \citep{TTV}. However, such an analysis would be much more robust and
informative if the planetary orbits are additionally constrained by radial velocities.
\item New physical effects may be revealed for some very close-in planets, owing to the
tidal interaction with the host star. We refer here to the case of WASP-12~b that
demonstrated apparently decaying orbital period, as derived from transit times
\citep{Maciejewski16}. But physical interpretation of such an effect is not unique: the
planet may either spiral down onto the star indeed, or it may udergo a tidal apsidal drift
\citep{Patra17}. Another possible explanation is that transit times are affected by a
variable light-travel time delay, owing to an unseen distant companion that induces a
long-period barycentric motion of the star and its transiting planet. The Doppler data are
very important in resolving such ambiguities. They may provide narrow constraints on the
orbital eccentricity, hence helping to distingush tidal apsidal precession from the true
orbit decay. From the other side, Doppler data may easily reveal the unseen companion if it
is indeed responsible for the apparent TTV effect, or they can rule out the existence of
such a companion.
\item The Rossiter-McLaughlin effect allows to determine whether the planet orbit is
aligned with the stellar equator or not \citep{GaudiWinn07}. This effect reveals itself
only in Doppler data, however its magnitude and shape depends on the stellar limb-darkening
law and average characteristics of the stellar spectrum, see e.g.
\citep{BaluevShaidulin15}. Transit observations would yield estimations of the
limb-darkening coefficients that are necessary for accurate modelling of the
Rossiter-McLaughlin effect.
\item Simultaneous RV and transit observations of the same star might be useful to detrend
activity effects and thus improve the accuracy of the analysis, see e.g. \citep{Aigrain12}.
\end{enumerate}

After the exoplanetary era started in 1990s thanks to 51~Pegasi discovery
\citep{MayorQueloz95}, the number of known exoplanetary candidates has now grown to several
thousands (see \emph{The Extrasolar Planets Encyclopaedia} database maintained by
\citet{Schneider11}). Before the Kepler space mission (\texttt{kepler.nasa.gov}),
ground-based RV surveys clearly overperformed the transit ones, so the Doppler technique
provided the main contribution in the detection statistics. Presently, the majority of
exoplanetary candidates are owed to the Kepler transit programme, but many of these Kepler
detections still remain insufficiently reliable. This highlights once again that both the
RV and transit method remain somewhat deficient, if they are used alone.

A variety of computer software is available to facilitate exoplanets detection and
characterization, based on the radial-velocity or transit data, or both
\citep{Meschiari09,Eastman13,Barragan17a,Barragan17b}. Multiple other works focus more on
the theory and the associated analysis methods, rather than software implementation
\citep{WrightHoward09,Pal10,Hara17}. See \citet{TRVTools} for a complete review of over
$40$ software tools available.

In this context, the primary goal of \PlP is to join the both approaches. The intention is
to host a wide set data analysis tools under the same umbrella, so that this toolbox could
be used solely alone if necessary, from the beginning to the end of the exoplanetary
analysis. Simultaneously, those methods are not just the textbook ones: most of them were
worked out by us over the past decade. All these algorithms have passed an extensive
real-world testing on practical exoplanetary cases. See the references in \citep{Baluev13a}
and below.

In the further sections, we discuss the new \PlP functionality introduced in its
version~3.0, along with the associated theory. This paper does not say anything about the
practical use of \PlP, its internal data organization, etc. The necessary Technical Manual
is provided in a standalone PDF file downloadable together with the \PlP sources.

\PlP source is available for download at the URL
\texttt{http://sourceforge.net/projects/planetpack}.

\section{Enhanced noise modelling}
\subsection{Some GP noise models typically used in practice}
\label{sec_bmod}
On their way to the detection of an Earth-twin exoplanet, Doppler programmes faced the need
to remove effects of stellar activity from the measurements \citep{Fischer16}. One popular
technique to partially bypass this ``activity barrier'' is to model the Doppler noise by
Gaussian Processes, or GPs \citep{GPML}. This method was adopted in multiple works already,
see e.g. \citep{Baluev11,Baluev13a,Baluev13c,FerozHobson14,AngladaEscude14,Rajpaul15}. In a
similar way, GPs can be used with photometric data, either to model the red noise in a
transit lightcurve \citep{Baluev15a,Barclay15}, or to characterize stellar rotation periods
\citep{Angus18}.

Presently, GPs can be deemed as one of standard noise modelling techniques in the field of
exoplanets detection \citep{Nelson18}, as well as in the general time-domain astronomy
\citep{Foreman-Mackey17}.

Let the noise ${\cal N}(t)$ be a centered GP. Then it is solely defined by its covariance
function:
\begin{equation}
\kappa(t_1,t_2) = \cov({\cal N}(t_1),{\cal N}(t_2)) = \expect \left({\cal N}(t_1) {\cal N}(t_2)\right).
\label{covdef}
\end{equation}
Based on $\kappa$, we can also introduce the variance $d(t)$ and the correlation function
$\rho(t_1,t_2)$:
\begin{equation}
d(t)=\var({\cal N}(t)) = \kappa(t,t), \quad \rho(t_1,t_2) = \frac{\kappa(t_1,t_2)}{\sqrt{d(t_1) d(t_2)}}.
\end{equation}

Quite often, ${\cal N}(t)$ is assumed stationary, which imples that its autocorrelation is
actually a function of just the time lag $\Delta t=|t_2-t_1|$, or $\kappa=\kappa(\Delta
t)$. In such a case the power spectrum $P(\omega)$ of the random process ${\cal N}(t)$ can
be expressed by means of the Wiener-Khinchin (WK) theorem via the Fourier transform of the
autocorrelation:
\begin{equation}
 P(\omega) = \hat\kappa(\omega) = \int\limits_{-\infty}^{\infty} \kappa(t) \exp(i\omega t) dt
\label{powsp}
\end{equation}
It is important that $P(\omega)$ cannot be negative, by its definition. Hence, the WK
theorem can serve as a test of physical admissibility of the correlation model. Assuming
some model $\kappa(\Delta t)$, we should substitute it to~(\ref{powsp}), and verify that
the resulting $P(\omega)$ never turns negative. If this test is failed then the selected
$\kappa(\Delta t)$ cannot describe any physical random process. Note that the WK theorem
itself does not require the random process to be necessarily Gaussian, so this test remains
valid even if $\mathcal N(t)$ has some deviation from strict normality.

Now, let us briefly consider several frequent choices of the GP model.

The first and the most primitive noise model is, of course, the white noise ${\cal W}(t)$.
Its autocorrelation is expressed by the Dirac delta function:
\begin{equation}
\cov({\cal W}(t_1),{\cal W}(t_2)) = \delta(t_2-t_1), \quad P(\omega) \equiv 1.
\label{covW}
\end{equation}
This definition implies that the variance of $\var{\cal W}(t)$ is infinite. It is possible
to select another normalization by using the Kronecker delta $\delta_{tt'}$ instead of
$\delta(t-t')$, thus making the variance finite, but this would imply another degeneracy,
$P(\omega)\equiv 0$. This actually means that the white noise is basically a degenerate
mathematical abstraction that does not exist in the real physical world. It can serve only
as a very basic approximation of the data.

A few models were used by \citet{Baluev11,Baluev13a} for the so-called ``red'', or
low-frequency, noise in Doppler data:
\begin{equation}
\cov({\cal R}(t_1),{\cal R}(t_2)) = \left\{ \begin{array}{l}
\exp\left(-\frac{|t_2-t_1|}{\tau}\right),\\
\exp\left[-\frac{(t_2-t_1)^2}{2\tau^2}\right],\\
\frac{1}{1+(t_2-t_1)^2/\tau^2}.
\end{array} \right.
\label{covR}
\end{equation}
The second of these models was also suggested by \citet{Rajpaul15}, also for the use with
Doppler data.

The covariances~(\ref{covR}) imply the following power spectra:
\begin{equation}
P(\omega) = \left\{ \begin{array}{l}
\frac{2\tau}{1+\tau^2\omega^2},\\
\sqrt{2\pi}\, \tau \exp\left(-\frac{\tau^2\omega^2}{2}\right),\\
\pi\tau \exp(-\tau |\omega|).
\end{array} \right.
\label{powR}
\end{equation}
All them have an unimodal bell-like shape centered at $\omega=0$.

Now let us start approaching the problem from another direction. Consider the sinusoidal
variation:
\begin{equation}
{\cal H}(t) = a \cos\omega_0 t + b \sin\omega_0 t,
\end{equation}
where $\omega_0$ is a fixed frequency, while $a$ and $b$ are independent random quantities
obeying standard normal distribution (mean zero, variance unit). Since $a$ and $b$ are
random, each value of ${\cal H}(t)$ is also random. It follows the same standard Gaussian
distribution, with zero mean and constant variance $d(t)\equiv 1$, while any set $\{{\cal
H}(t_k)\}$ is a multivariate Gaussian vector. Therefore, ${\cal H}(t)$ is a stationary GP.
Its covariance function is:
\begin{equation}
\cov({\cal H}(t_1),{\cal H}(t_2)) = \cos\left[\omega_0 (t_2-t_1)\right],
\label{covH}
\end{equation}
and the power spectrum:
\begin{equation}
P(\omega)=\frac{1}{2}\left[ \delta(\omega-\omega_0) + \delta(\omega+\omega_0) \right].
\label{powH}
\end{equation}

Simultaneously, each instance of $\cal H$ is just a sinusoid. As such, we can say
that~(\ref{covH}) represents the covariance function of a sinusoid. And ${\cal H}(t)$ is
often called the ``harmonic GP''.

Thanks to the strict periodicity of the covariance~(\ref{covH}), each instance of ${\cal
H}(t)$ is a periodic function. Even though we defined ${\cal H}$ as a random process, by
looking at any its single instance it would not be possible to note typical signs of a
random variation (a noise). In other words, instances of such a process always look like a
deterministic function (though its phase and amplitude depends on the particular instance).
This appears because~(\ref{covH}) does not decay for large $\Delta t$, so even distant
values of ${\cal H}(t)$ are strictly binded with each other, while the phase of the
variation is presereved over time.

A similar behaviour occurs for any other random process ${\cal P}(t)$ that has a strictly
periodic covariance function $\kappa(\Delta t)$, for example for
\begin{align}
\cov({\cal P}(t_1),{\cal P}(t_2)) &= \exp\left\{-\frac{2}{\lambda^2}\sin^2\left[\frac{\pi}{P}(t_2-t_1)\right]\right\} \nonumber\\
 &\propto \exp\left\{\frac{1}{\lambda^2}\cos\left[\frac{2\pi}{P}(t_2-t_1)\right]\right\},
\label{covP}
\end{align}
from \citep{Olspert18}, or for any other $P$-periodic $\kappa(\Delta t)$. Each instance of
such a random process is a strictly periodic variation, though not necessarily sinusoidal.
The parameter $\lambda$ in~(\ref{covP}) controls the shape of the resulting periodic
variation.

In general, the power spectrum of such a periodic random process is, by analogy
with~(\ref{powH}), a sum of multiple delta impulses centered at the discrete frequencies
$\omega_n = 2\pi n/P$, where $n=\pm 1,\pm 2,\ldots$ It is required that all these impulses
are positive, or otherwise the selected model $\kappa(\Delta t)$ would become non-physical.

The natural further step is to construct a mixture of an autocorrelated noise with a
decaying $\kappa_{\rm d}(\Delta t)$, like one of~(\ref{covR}), and of a periodic model
$\kappa_{\rm p}(\Delta t)$, like~(\ref{covH}) or~(\ref{covP}). This can be done by
constructing the product of the covariances:
\begin{equation}
\kappa(\Delta t) = \kappa_{\rm d}(\Delta t) \kappa_{\rm p}(\Delta t).
\end{equation}
This random process would carry joint signs of a random noise and of a periodic variation,
i.e. it is a quasiperiodic GP. The power spectrum of such a quasiperiodic noise represents
a convolution of the parent power spectra $P_{\rm d}(\omega)$ and $P_{\rm p}(\omega)$. This
results in a discrete sequence of peaks in $P(\omega)$, owed to the periodic part $P_{\rm
p}$, and each such peak has the same bell-like shape (a broadened delta function), which is
determined by the noisy part $P_{\rm d}$.

One of the most simple choices is to combine the exponential red noise ${\cal R}(t)$ with
the harmonic oscillation ${\cal H}(t)$:
\begin{equation}
\kappa(\Delta t) = {\text e}^{-\beta|\Delta t|} \cos\omega_0\Delta t,
\label{covDS}
\end{equation}
which yields the covariance in the form of a decaying sinusoid. It is similar to the ones
implemented by \citet{Foreman-Mackey17}.

However, \citet{Rajpaul15} suggested a bit more complicated model, which was based
on~(\ref{covP}) and on the squared-exponential red noise from~(\ref{covR}):
\begin{equation}
 \kappa(\Delta t) = \exp\left\{-\frac{\Delta t^2}{2\tau^2} - \frac{2}{\lambda^2}\sin^2 \frac{\pi\Delta t}{P} \right\}.
\label{covQP}
\end{equation}

Both the models~(\ref{covDS}) and~(\ref{covQP}) are valid covariance functions that can
describe a quasiperiodic random process. Heuristic models of these types represent a good
``workhorses'' that allow to adequately handle e.g. Doppler or photometric noise appearing
due to the stellar activity (e.g. spots), coupled with stellar rotation. The quasiperiodic
model by \citet{Rajpaul15} is perhaps a bit more general, thanks to an additional tuning
parameter $\lambda$ that affects the non-sinusoidal shape of the quasiperiodic
variation.\footnote{Note that we changed some of the original designations by
\citet{Rajpaul15} to adapt them to our notation system.}

Yet another possible direction of generalizing is to consider non-stationary GPs, when the
covariance $\kappa(t_1,t_2)$ cannot be reduced to a single argument. Such an attempt was
made in \citep{Baluev14a} to analyse long-term activity variations in 55~Cnc. We used the
following model in that work:
\begin{align}
\kappa(t_1,t_2) &= {\text e}^{-|t_2-t_1|} \cos\frac{\Omega t_1+\lambda}{2} \cos\frac{\Omega t_2+\lambda}{2} \nonumber\\
 &= \frac{1}{2} {\text e}^{-|t_2-t_1|} \left[\cos \Omega\frac{t_2-t_1}{2} + \cos\left(\Omega \frac{t_1+t_2}{2} + \lambda \right) \right].
\label{covNS}
\end{align}
The intention was to construct a mathematical model with a periodic modulational factor
that could describe long-term activity variation, and simultaneously keep the formula
symmetric with respect to swapping $t_1$ and $t_2$. In particular, the variance for this
model varies sinusoidally:
\begin{equation}
d(t) = \frac{1}{2} \left[1 + \cos(\Omega t + \lambda) \right].
\end{equation}

As we can see, there is a plenty of mathematical GP models that could explain, more or less
adequately, some specific noise effects. However, their common concern is that all these
models are heuristic and are highly arbitrary. The GP models used in the astronomical
practice typically lack physical motivation of their choice. They are often chosen based on
either their popularity or just mathematical simplicity. They do not rely on physical
models of the star and appear just mathematical toy models instead.

An exception is the work by \citet{Foreman-Mackey17}, where the authors give some
explanation of their GP kernels, based on the physical model of a dumped harmonic
oscillator (DHO).

It is still impossible to construct a feasible physically justified model of the stellar
noise in Doppler or photometric data. But nevertheless we may try to take into account some
very basic and simple principles of physical self-consistency, or at least to investigate
how a particular assumption concerning the red noise generating mechanisms may affect our
models. Such an attempt is presented in App.~\ref{sec_cause}, where we adopt one such
self-consistent view on the GPs that can probably cover a wider range of models than just
the DHO concept. We consider this self-consistent framework as possible basis to construct
GP models used in \PlP.

\subsection{Elementary GP ``bricks'' accessible in \PlP}
\label{sec_gpnoise}
\PlP~3.0 allows to build a multicomponent noise model from multiple ``GP primitives''. It
is assumed that the noise is a sum of statistically independent GP contributions, and each
has a simple form of the covariance function $\kappa$. Then the cumulative covariance
function is plainly a sum of these elementary ones.

Current version 3.0 of \PlP allows to combine the following stationary ``GP primitives'':
the white noise~(\ref{covW}), and the three types of the red noise~(\ref{covR}).

Each stationary component, either white or red, can be promoted to a nonstationary version
by means of a sinusoidal modulation:
\begin{align}
\kappa_{\rm nonstat}(t_1,t_2) &= d(\min(t_1,t_2))\, \kappa_{\rm stat}(\Delta t), \nonumber\\
d(t) &= D_0 + D_m \cos(\Omega t + \lambda).
\label{vcmod}
\end{align}
Such nonstationary GPs can be used to model e.g. long-term stellar activity that may
modulate short-term noise characteristics \citep{Baluev14a}. The formula~(\ref{vcmod}) is a
bit different from the nonstationary model~(\ref{covNS}). The new formula is based on the
``adiabatic'' approximation~(\ref{varcovmod}) motivated by our self-consistent GP
construction method given in the App.~\ref{sec_cause}. The modulation is ``adiabatic'' in
the sense that $\Omega$ is assumed much smaller than the red-noise decay parameter
$\beta=1/\tau$. This corresponds to a relatively slow modulation effect like e.g. the
magnetic activity cycle. We note, however, that in this adiabatic case our new
model~(\ref{vcmod}) appears close to the old one~(\ref{covNS}), so the results of the
analysis should not change too much.

A single white-noise term in the compound model is mandatory, while the rest is an
arbitrary combination of autocorrelated terms:
\begin{equation}
\kappa(t_1,t_2) = \kappa_{\rm WN}(t_1,t_2) + \sum_{k=1}^{n} \kappa_{{\rm RN},k}(t_1,t_2).
\end{equation}
Each term in this model is parametrized by its own set of free variables: the magnitude of
the jitter $\sigma_\star$, the red-noise correlation timescale $\tau$ (if this is a
red-noise term), optionally the three modulation parameters: $\sigma_{\star,\rm
mod}=\sqrt{D_m}$, $P_{\rm mod}=2\pi/\Omega$, and $\lambda_{\rm mod}$. These parameters can
be fitted all independently, or fixed or mutually binded as desired (this option is
inherited from the \PlP~2.0 constrained fitting mechanism).

Three types of the white noise model $\kappa_{\rm WN}$ can be used: the multiplicative, the
additive with trucation, and the regularized modification of the additive model. The latter
one is now used by default thanks to its high robustness confirmed by practical
computations, see \citep{Baluev14a,Baluev15a} and the forthcoming work (Baluev et al.,
2018, in prep.). These noise models are described detailedly in \citep{Baluev14a} and in
the \PlP Technical manual.

In the further versions of the 3.x series we plan to add support for quasiperiodic noise
models, like~(\ref{covDS}) and~(\ref{covQP}), and possibly the phase-shifted
version~(\ref{varcovexpds}), after they have enough testing. These GP kernels can be used
to model stellar rotation effect.

The fitting of a GP model is performed using generally the same maximum-likelihood
framework as in the 1.x \PlP series. Some details of the theory and numeric calculation are
given in \citep{Baluev13a}. The inversion of the $N\times N$ data covariance matrix is
achieved through the Cholesky decomposition, which is numerically stable and quick relative
to other methods ($\sim N^3/6$ multiply operations on a dense matrix). To improve the
computation speed, the covariance matrix is first made sparse by means of forcing all small
correlations ($|\rho(\Delta t)|<\varepsilon$) to exact zero. This is done in a smooth
manner in order to avoid undesired numerical effects that could appear due to artificial
discontinuties in $\rho(\Delta t)$. Then the covariance matrix becomes a symmetric band
matrix, i.e. only $2K+1$ central diagonals remain non-zero, where $K\ll N$ usually. In such
a way \PlP uses dedicated linear algebra routines that profit from the band matrix
structure. Such routines are available in the {\sc OpenBLAS} library, for instance. The
Cholesky decomposition algorithm, which is not present in {\sc OpenBLAS}, was also
programmed to take into account the banded structure of the input matrix, resulting in just
$\mathcal O(NK^2)$ arithmetic operations.

\section{Transit fitting with \PlP}
\label{sec_transit}
\subsection{The motivation}
The transit fitting was introduced in \PlP~2.x series as an experimental standalone
pipeline processing lightcurve data (without radial velocities). This pipeline had a
relatively narrow purpuse and was run by just a single \PlP command \texttt{transitfit}.
Accepting an input set of lightcurves, each with a complete or partial planetary transit,
the pipeline derived their transit timing data, for further investigation of possible
transit timing variations (TTVs).

In v.~3.0 we provide a more powerful toolbox for self-consistent fitting of the transit
data, with or without radial velocities (see sect.~\ref{sec_joint}). However, the TTV-only
fitting pipeline is preserved, as its goal is important, albeit more narrow. This pipeline
is now tested well enough and was improved in several aspects.

The basics of the transit curve modelling were already presented by \citet{Baluev15a}, and
more details soon appear in the forthcoming update of this paper (Baluev et al., 2018, in
prep). Here we do not replicate all their formulae. We only highlight the main points and
improvements introduced with \PlP~3.0.

\subsection{The models}
The transit models remained basically the same as in \citep{Baluev15a}. This includes a
circular curved orbital motion of the planet (i.e., assuming zero eccentricity) and the
quadratic model for the limb-darkening. The model of the magnitude drop itself is
mathematically equivalent to the classic one presented by \citet{MandelAgol02} and to the
one by \citet{AbubGost13}, although technically we use the formulae by
\citet{BaluevShaidulin15} that put photometric and spectroscopic transits in the same
modelling framework.

Our transit model has $4$ kinematic parameters: (i) the mid-time of the transit $t_c$, (ii)
the half-duration of the transit $t_d$, defined as $1/2$ of the time spent between the
first and fourth contacts, (iii) the impact parameter $b$, measuring the smallest projected
separation between the planet and star centers (divided by the star radius), and (iv) the
projected planet/star radii ratio $r$ that simultaneously determines the transit depth and
the duration of the ingress/egress phases.

We assume the quadratic limb-darkening model with two coefficients to be determined, $A$
and $B$. The brightness of a point on the visible stellar disc, observed at a given
separation from its center, $\rho$, is modelled as
\begin{equation}
I(\rho) = 1 - A (1-\mu) - B (1-\mu)^2, \quad \mu=\sqrt{1-\rho^2}.
\label{bright}
\end{equation}

Sometimes the model~(\ref{bright}) becomes ill-fitted because of poor data quality, and
then it can turn non-physical, owed to bad values of $A$ and $B$. To avoid this issue, \PlP
includes internally the following mandatory constraints on these coefficients:
\begin{equation}
A+B\leq 1, \quad A+2B\geq 0, \quad A \geq 0.
\label{ldcon}
\end{equation}
According to \citet{Kipping13,Baluev15a}, these constraints are necessary to have $I(\rho)$
always positive and monotonically decreasing (any ``limb brightening'' is disallowed).

Technically, the constrained fitting honouring~(\ref{ldcon}) is performed in an implicit
manner, by using the following trigonometric replacement:
\begin{equation}
A = (1-\cos\varphi) \sin^2\theta, \quad B = \cos\varphi \sin^2\theta.
\end{equation}
Internally, \PlP uses auxiliary variables $\theta$ and $\varphi$ as primary fittable
parameters. Whatever real values they attain, the resulting values of $A$ and $B$ always
satisfy~(\ref{ldcon}). From the other side, each point $(A,B)$ in the domain~(\ref{ldcon})
maps to some real-valued $(\theta,\varphi)$:
\begin{equation}
\sin^2\theta = A+B, \quad \cos\varphi = \frac{B}{A+B}.
\label{angAB}
\end{equation}
Note that from~(\ref{ldcon}) it follows that $A+B \geq |B|$, meaning that $A+B$ is never
negative (though $B$ can become negative). The trigonometric parametrization~(\ref{angAB})
is alternative to the $(q_1,q_2)$ one proposed by \citet{Kipping13}: $q_1=\sin^4\theta$,
$q_2=\sin^2\varphi$. But the parameters~(\ref{angAB}) allow us to avoid dealing with
boundaries. Every point in the $(\theta,\varphi)$ plane corresponds to some meaningful
limb-darkening coefficients.

The lightcurve model may optionally include a polynomial trend with fittable coefficients.
Such a trend is necessary to take into account various drifting effects, e.g. the effect of
airmass or other types of systematic variations that appear frequently in the transit data.
Each transit lightcurve may have an individual fittable trend with a separate set of trend
coefficients (although their polynomial order must be the same). We find that cubic trends
represent a good compromise between the model adequacy and its parametric complexity. \PlP
is currently not capable to detrend the data against the airmass function or other
additional indicators, but this might be a work for future development.

\PlP may optionally include a single fittable red-noise noise term with an exponentially
decaying correlation function~(\ref{covR}). This can be done either for all lightcurves or
only for some selected ones. In any case, the red noise in different lightcurves is fitted
independently. It is also possible to autodetect red noise in the input photometry (see
below).

The fitting of these models is performed by means of the maximum-likelihood approach with a
preventive bias correction in the noise jitter. This method remained practically unchanged
from \PlP~1.0, and is explained in details in
\citep{Baluev08b,Baluev11,Baluev13a,Baluev13c}.

\subsection{The pipeline}
The transit fitting pipeline include the following steps that invlove a sequential increase
of the model complexity:
\begin{enumerate}
\item Perform a preliminary fit constraining the mid-times on a regular grid (linear ephemiris with
free time shift and stride); fixing the impact parameter at an intermediary value ($b\simeq
1/\sqrt 2$), and fixing the limb darkening coefficients at $A=B=0.25$ (or $\theta=\pi/4$
and $\varphi=\pi/3$).
\item Refit after releasing the mid-times and the impact parameter, but still holding the
limb darkening coefficients fixed.
\item Refit after releasing the limb darkening coefficients.
\item Refit after a releasing the limb darkening coefficients for the best-quality
lightcurves (those with r.m.s. $<0.05$ of the transit depth and complete transits).
\end{enumerate}
This pipeline remains generally the same as used in \citep{Baluev15a} and \PlP~2.0. The
main changes are in parametric constraints. In the 2.x series the following parameters we
mutually binded (across different lightcurves): impact parameter $b$, transit half-duration
$t_d$, the planet/star radii ratio $r$, and the limb-darkening parameters $A$ and $B$. Only
transit timings were fitted separately, allowing for further analysis of transit timing
variation (TTV), but not of possible transit duration variation (TDV), or variations of the
apparent transit depth (via $r$). Also, it was impossible to fit the limb-darkening
coefficients separately for lightcurves obtained in different photometric bands.

In the new \PlP~3.0 these restrictions are not mandatory and can be removed if desired.
Most importantly, the limb-darkening coefficients can be fitted separately on the
band-by-band basis, i.e. it is possible to bind $A$ and $B$ only between those lightcurves
that were obtained in the same (or similar) photometric filter. See Technical
manual for the details.

This \PlP functionality allows for an interesting by-product research: investigate an
experimental dependence of $A$ and $B$ from the spectral band, and then compare these
estimations with theoretic predictions (Baluev et al., 2018, in prep.).

Already \PlP~2.0 allowed to autodetect red noise in the input photometry. In such a case,
only those lightcurves would gain a red-noise model term, in which this red noise appeared
statistically significant and not ill-fitted. However, this algorithm was very
time-consuming, because it ``probed'' each lightcurve individually, i.e. tried to fit it
with and without the red-noise term and then compared the likelihood of these fits. The
fitting of the correlated noise is computationally hard in itself (due to inversion of
large covariance matrices), especially if such a fit need to be re-run multiple times.

In v.~3.0 we significantly improved the speed of this computation, by using fast linear
algebra libraries (see sect.~\ref{sec_comput}), and by optimizing the order in which the
individual lightcurves are probed.

The main principle of this optimized sequence: process good data last. The motivation is
that low-quality data usually do not demonstrate any detectable or robustly-fittable red
noise. Therefore, it is reasonable to probe them first, just to quickly ensure that their
red noise is ill-fitted, and then continue to fit them with the fast white noise model
while probing the red noise in the remaining high-quality data. If instead the high-quality
data are processed first, then we likely detect some robust red noise in them, thus slowing
all further computation down at the very beginning stage of the analysis.

To identify transit data of a higher quality, the following ``quality characteristic'' of a
lightcurve is computed:
\begin{equation}
\mathcal Q_{\rm lc} = \mathcal Q \sqrt{2 t_{\rm d}}, \quad \mathcal Q = \frac{\sqrt{\text{measurements density}}}{\text{residuals r.m.s.}}
\end{equation}
The quantity $1/\mathcal Q$ determines the uncertainty offered by a ``standard'' chunk of
the lightcurve of a unit length. The uncertainty of an arbitrary chunk of length $t$ scales
as $1/(\mathcal Q\sqrt t)$, so $\mathcal Q \sqrt t$ can be accepted as a rough ``quality
characteristic'' of the chunk, while $\mathcal Q \sqrt{2t_{\rm d}}$ represents the quality
characteristic of the in-transit portion of the lighcurve. \footnote{In this rough and
indicative characteristic we neglect possible red noise, so even neighbouring measurements
are assumed uncorrelated.}

\subsection{The code}
We no longer rely on the code by \citet{AbubGost13}. Based on their formalism, we
constructed a more general theory that incorporates models of the transit curve and of the
Rossiter-McLaughlin effect in the same self-consistent framework \citep{BaluevShaidulin15}.
\PlP~3.0 relies on that more general theory and uses its own computing code. The elliptic
functions, required for these computations, are computed using the algorithms developed by
\citet{Fukushima13}.

\section{Joint Doppler+transit fits with \PlP}
\label{sec_joint}
Probably the most useful feature of the new \PlP~3.0 is the joint self-consistent fitting
of the RV and transit data. In itself such a task is not novel, see e.g. {\sc EXOFAST}
software by \citet{Eastman13} and {\sc pyaneti} by \citet{Barragan17b}, but in the context
of the \PlP this means that all dedicated functionaly inherited from the legacy RV-only
\PlP now becomes available for such mixed fits too. This includes: $N$-body fitting,
complicated noise models discussed above (both for the RV and for the photometry),
periodograms, constrained fitting, statistical model comparisons and simulations
\citep{Baluev13c}. The self-consistency is achieved because when performing such a joint
fit, \PlP builds a 3D model of the entire planetary system, computing the motion of all
planets in accordance with either $N$-body integration or multi-Keplerian formulae. Such an
analysis is more accurate than the TTV fitting pipeline discussed above.

Using these tools it is possible to solve quite complicated tasks, for example to seek a
planet that induced the observed TTV in mid-transit times of the known planet. Thanks to
the self-consistency, this can be done in a periodogram-like manner, and avoiding any work
with intermediate-stage TTV data.\footnote{This can be achieved by varying the orbital
period of the putative second planet, and optionally other its parameters, via the
\texttt{afit} command, and assuming the $N$-body modelling framework.}

\PlP~3.0 enables the user to fit the Rossiter-McLaughlin effect using the models presented by
\citet{BaluevShaidulin15}. This allows to estimate the stellar rotation parameters $v\sin
i$ and $\lambda$, optionally taking into account the limb-darkening coefficients and/or
correction coefficients $\nu$ and $\mu$ that depend on the average spectrum characteristics
\citep{BaluevShaidulin15}.

Finally, in v.~3.0 we added the possibility to fit dissipative orbital effects. This
includes the long-term period decay (e.g. due to tidal interaction with the star) and the
secular apsidal drift (also can appear in tidal interactions). The need for such models is
motivated by the unique case of the planet WASP-12~b that demonstrated clear TTV hints of
such non-Keplerian effects \citep{Maciejewski16,Patra17}.

The period decay is modelled linearly in terms of the planet mean-motion $n$ and
quadratically in terms of the orbital mean longitude $l$:
\begin{equation}
\frac{n(t)}{n_0} = 1 + \frac{t-t_0}{T_{\rm d}}, \quad l(t) = l_0 + n_0 (t-t_0) + \frac{n_0}{2T_{\rm d}} (t-t_0)^2
\end{equation}
Formally, in this model the planet would never fall on the star, but in practice we of
course consider very slow effect, $\dot n/n_0\ll 1$, and hence only very short piece of its
evolution, relatively to $T_{\rm d}$. In this case, the apparent orbital period would
decrease as
\begin{equation}
\frac{P(t)}{P_0} \simeq 1 - \frac{t-t_0}{T_{\rm d}},
\end{equation}
implying that $T_{\rm d}$ has the meaning of the remaining linear lifetime of the planet.
This $T_{\rm d}$ serves as a fittable parameter of the model (if this effect is turned on).

The effect of an apsidal drift is modelled as a linear change of the pericenter argument
$\omega$ (not to be mixed with the frequency argument $\omega$ appearing in
sect.~\ref{sec_gpnoise}):
\begin{equation}
\omega = \omega_0 + \frac{2\pi}{P_\omega} (t-t_0).
\end{equation}
Here, $P_\omega$ is the secular period of the apsidal revolution, and it serves as a
fittable parameter of the effect. \PlP allows to turn both the $T_{\rm d}$-,
$P_\omega$-effects on, or just one of them, or use the pure Keplerian model. These effects
are not available in $N$-body fits.

We recognize that several subtle but in some cases potentially important effects are not
yet implemented in the current \PlP code, but remain in our plans for future. This includes
secondary planetary eclipses, the effect of time shift between the transit and Doppler
data, appearing due to the light-travel delay,\footnote{The Doppler information is
imprinted when the light is emitted by the stellar surface, but the transit effect appears
a few seconds later, when this light is absorbed by the dark planet.}, simultaneous
multiple transits with interplanetary eclipse phenomenta,\footnote{Currently, \PlP allows
to fit multiple planets transiting the star, even simultaneously, but it assumes that the
cumulative magnitude drop is just the sum of individual planetary contributions, i.e. that
planets never eclipse each other.} and planet oblateness effect.

Another direction of future development is to include the analysis of astrometric data,
which will become especially important in the near future, after the exoplanetary results
from GAIA \citep{Gaia17,Gaia18} are released. Then \PlP should gain the ability to perform
the self-consistent fits using data of three types: Doppler, photometry, and astrometry.
This would enable the most exhaustive orbital characterization of exoplanetary systems.

\section{Improved computation performance via
{\sc BLAS} library and multithreading}
\label{sec_comput}
\PlP~3.0 is considerably more fast than its legacy versions. This was achieved by (i)
multithreading tools of the {C++11} language standard and (ii) migrating the most heavy
linear algebra to the CPU-optimized {\sc OpenBLAS} library. The multithreading was utilized
in \PlP~2.x series already, but the use of {\sc OpenBLAS} library is a new feature.
According to our benchmarks, this allowed to improve the computational speed by the factor
of $3$ in some cases. The effect is especially noticable when analysing large datasets,
$N\sim 10^3$, which is typical for photometry. In such a case most computation time is
spend in large matrix-matrix multiplications (if the noise is white). Also, {\sc OpenBLAS}
allows for some partial multithreading of the linear algebra operations. This is used in
\PlP whenever the selected analysis algorithm does not allow easy parallelization in
itself.

The use of {\sc OpenBLAS} is not mandatory. \PlP can be compiled without {\sc OpenBLAS}, if
it is not available on a given computer. Then it will rely on its own implementation of the
necessary BLAS routines, but this means no profit from highly-optimized libraries. In some
future, we consider a possibility to build \PlP with any BLAS library chosen by the user.
However, many of the available BLAS implementations do not support multithreading, so we
currently stopped on {\sc OpenBLAS} that does. But another interesting choice might be the
GPU-based {\sc cuBLAS}.

\section{Conclusions and plans for further development}
\label{sec_conc}
\begin{figure*}[!t]
\includegraphics[width=\textwidth]{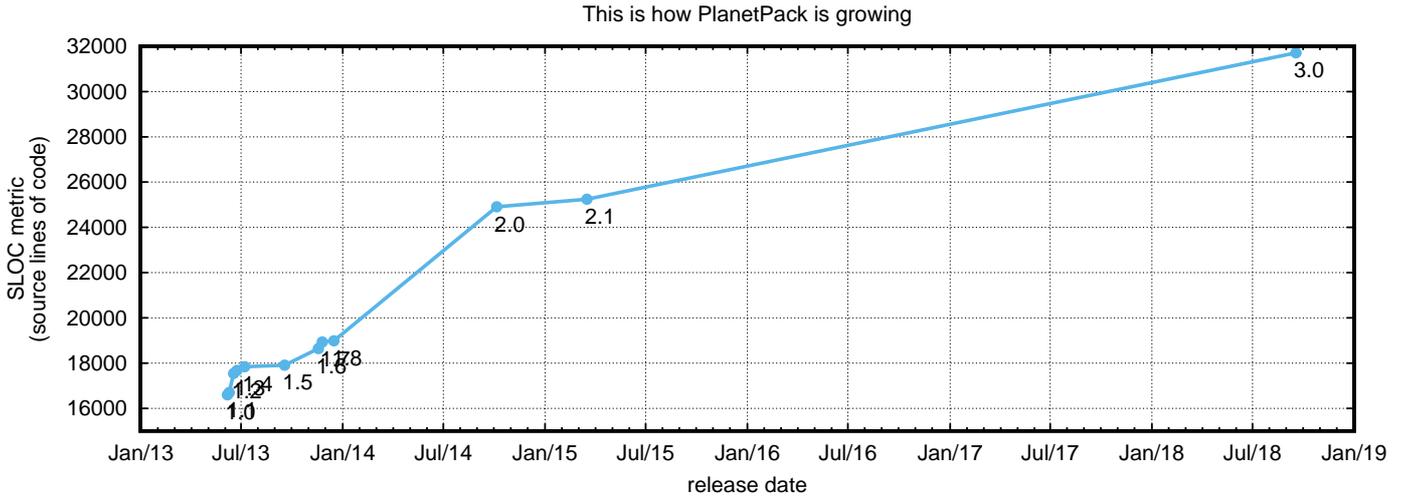}
\caption{Evolution of the \PlP source code.}
\label{fig_lines}
\end{figure*}

Since its first release in 2013, \PlP functionality grew significantly, owing to bug fixes
as well as to new analysis algorithms. An approximate impression of \PlP code evolution can
be obtained from Fig.~\ref{fig_lines}. As we can see, there was a long pause beween the
last 2.x release and the new 3.0 one, but nonetheless there was a remarkable source code
expansion between them.

In the future, it would be useful to add capabilities of dealing with astrometric data,
because of the coming era of GAIA astrometry.

Astrometry is a pretty feasible technique of exoplanets detection, though the ability of
GAIA to reliably detect and characterize long-period ($P\gtrsim 10$~yr) exoplanets
currently looks possibly doubtful, because of the relatively $5$-year duration of the
mission. Contrary to space missions, ground-based programmes (Doppler or transit ones) are
able to accumulate data over much longer terms. Nonetheless, we consider the inclusion of
astrometric data analysis as a necessary condition for the next major release of \PlP.

Among more minor and technical things, the following might deserve implementation in \PlP:
\begin{enumerate}
\item Quasiperiodic noise like~(\ref{covQP}) or similar, first on the todo list.
\item Reducing stellar Doppler noise based on its correlation with activity indicators
\citep{Anglada-Escude12,AngladaEscude14}, and using similar detrending approach for
photometric data (against e.g. airmass).
\item Fitting of secondary planetary eclipses.
\item Improving speed from GPU computing, in particular by using the {\sc CUDA BLAS}
libraries in addition to CPU-based {\sc OpenBLAS}.
\end{enumerate}

Statistical methods implemented in \PlP rely on the frequentist treatment and
maximum-likelihood fitting with a preventive bias reduction in noise parameters. The
relevant theory was presented mainly in \citep{Baluev08b,Baluev13a}.

\PlP is a free and open-source software. We do not set any limitation on its use or on the
use of its source code, except for providing a proper reference to the present paper and
\citep{Baluev13c}.

\section*{Acknowledgements}
This work was supported by the Russian Foundation for Basic Research grant 17-02-00542~A
and by the Presidium of Russian Academy of Sciences programme P-28, subprogramme ``The
space: investigating fundamental processes and their interrelations''.

We acknowledge that a few of linear algebra algorithms used in \PlP represent re-worked
versions of the relevant GNU Scientific library subroutines; \PlP also relies on the
open-source {\sc OpenBLAS} and {\sc GNU readline} libraries.

We thank the anonymous reviewers of the manuscript for their fruitful comments and
suggestions.

\appendix

\section{A self-consistent framework to build Gaussian processes models}
\label{sec_cause}
\subsection{GP noise and the causality}
\label{sec_model}
Now, let us start from noticing that a correlated Gaussian noise process can be obtained by
integrating the white Gaussian noise. For example, the Wiener process (a classic model of
the Brownian motion) can be defined by integrating the standard Gaussian white noise ${\cal
W}(t)$:
\begin{equation}
{\cal B}(t) = \int\limits_0^t {\cal W}(t') dt'.
\label{BB}
\end{equation}
This stochastic integral yields us a Gaussian random process, as far as it is a linear
functional of $\cal W$, which was Gaussian. The covariance characteristics of $\cal B$ can
be obtained by interchanging the mathematical expectation operator in~(\ref{covdef}) with
the integration in~(\ref{BB}):
\begin{align}
\cov({\cal B}(t_1),{\cal B}(t_2)) &= \int\limits_0^{t_1} \int\limits_0^{t_2} \cov({\cal W}(t_1'),{\cal W}(t_2')) dt_1' dt_2' \nonumber\\
 &= \min(t_1,t_2), \nonumber\\
\var {\cal B}(t) &= \int\limits_0^t \int\limits_0^t \cov({\cal W}(t_1'),{\cal W}(t_2')) dt_1' dt_2' \nonumber\\
 &= t.
\end{align}
In particular, the variance of ${\cal B}(t)$ is constantly growing, while the correlation
function is expressed as $\rho(t_1,t_2) = \min\left(\sqrt{t_1/t_2},\sqrt{t_2/t_1}\right)$.

It is important that $B(t)$ satisfies certain causality principle: its values only depend
on the past values of the parent process ${\cal W}(t)$.

Due to the constantly growing variance, $B(t)$ does not suit our needs, however. We need
our noise model to be either strictly stationary or at least long-term bounded. The
nonstationarity of $B(t)$ appears because the integration in~(\ref{BB}) has a variable
upper limit $t$. To have more or less constant variance, we must limit the integration to a
time segment of a constant length.

Guided by this observation, we adopted the following weighted-integral model:
\begin{eqnarray}
x(t) &=& \int\limits_{-\infty}^t w(t-t') n(t') dt', \label{irn} \\
& & w(t)=0\quad {\rm for}\quad t<0, \label{cas}\\
& & w(t)\to 0\quad {\rm for}\quad t\to +\infty,\\
& & n(t) = \sqrt{D(t)} {\cal W}(t).
\end{eqnarray}
This formal definition can be given the following physical understanding. The observable
correlated process $x(t)$ is a cumulative sum of infinitesimal independent ``kicks'',
generated by the underlying white-noise ``activity'' process $n(t)$. Optionally, this
activity process may have a time-variable intensity defined by the function $D(t)$.
Additionally, there is also a memory (a.k.a. response) function $w(t)$ that defines how
much a past ``kick'' affects the current value of $x$.

The causality restriction settles the requirement~(\ref{cas}), meaning that future values
of the activity do not affect the current observable value. Also, the effect of the
activity should naturally decay with time, so that only recent ``kicks'' have a major
effect on the observable process.

It is important that definition~(\ref{irn}) is self-replicatable in the sense that by
applying this formula once again to the same process, we obtain an integral transformation
of the same type, but with different $w(t)$. Therefore, even if there are multiple physical
effects that ``soften'' $n(t)$ sequentially, the final result can be always represented in
the simple form~(\ref{irn}). Therefore, our assumption that $n(t)$ is white noise does not
reduce the generality of~(\ref{irn}) as much as it may seem. If $n(t)$ was a red noise
instead then this $n(t)$ would likely be generated by a softening mechanism of the same
type~(\ref{irn}), with some parent $n_1(t)$, which would be ``more white'' than $n(t)$. In
such a case we could join the double integration into one, assuming $n_1(t)$ be the actual
parent process. And so on, until we reach some white ``progenitor process'' $n_\infty(t)$.

The most important assumption hidden in~(\ref{irn}) is that all the softening effects are
linear, i.e. they can be represented by a linear integral transform.

The covariance characteristics of so-defined $x(t)$ are:
\begin{align}
\kappa(t_1,t_2) &= \int\limits_{-\infty}^{\min(t_1,t_2)} w(t_1-t') w(t_2-t') D(t') dt', \nonumber\\
d(t) &= \int\limits_{-\infty}^t w^2(t-t') D(t') dt'.
\label{varcov}
\end{align}
In particular, constant intensity $D(t)$ means that the activity process $n(t)$ is
stationary, implying the stationary $x(t)$:
\begin{equation}
\kappa(\Delta t) = \int\limits_0^{+\infty} w(t') w(|\Delta t|+t') dt', \quad
d(t) = \int\limits_0^{+\infty} w^2(t') dt'.
\label{varcovstat}
\end{equation}
The power spectrum of $x(t)$ is then obtained by the WK theorem:
\begin{equation}
P(\omega) = |\hat w(\omega)|^2.
\end{equation}

By selecting a variable $D(t)$ we may introduce a controllable nonstationarity to both
$n(t)$ and $x(t)$. For example, in case of an adiabatic nonstationarity, when $D(t)$ varies
much slower than $w(t)$, the effect is reduced to a modulation:
\begin{align}
\kappa(t_1,t_2) &\simeq D(\min(t_1,t_2)) \int\limits_0^{+\infty} w(t') w(|\Delta t|+t') dt', \nonumber\\
d(t) &\simeq D(t) \int\limits_0^{+\infty} w^2(t') dt'.
\label{varcovmod}
\end{align}

The choice of the memory and intensity functions still remains rather arbitrary, and is
governed mainly by mathematical simplicity considerations. For example, let us assume an
exponential model:
\begin{equation}
w(t) = \exp(-t),\quad t\geq 0.
\end{equation}
In this case and for constant $D(t)\equiv 1$ we have
\begin{equation}
\kappa(\Delta t) = \frac{1}{2} \exp(-|\Delta t|), \quad
d(t) = \frac{1}{2},
\label{varcovexpstat}
\end{equation}
This implies the exponential correlation function already discussed above.

Other types of red noise from~(\ref{covR}) can be modelled by means of the integral
representation~(\ref{irn}). This can be achieved by taking square root of the corresponding
power spectrum~(\ref{powR}) and setting the weight function such that $\hat w(\omega) =
\sqrt{P(\omega)}$. However, it is important that the resulting $w(t)$ may violate the
causality restriction~(\ref{cas}). For example, for the Gaussian-shaped covariance,
$\kappa(t)=\exp(-t^2/2)$, the natural $w(t)$ is proportional to $\exp(-t^2)$, which is
symmetric with respect to the past and future.

The requirement of smoothness, that governed \citet{Rajpaul15} to select a
square-exponential model in~(\ref{covR}), does not seem physically necessary, as it might
appear incompatible with our causality restriction. Let us compute the derivative of the
covariance function $\kappa(\Delta t)$ from~(\ref{varcovstat}) at $\Delta t=0$:
\begin{equation}
\kappa'(\pm 0) = \pm \int\limits_0^{+\infty} w(t) w'(t) dt = \pm w^2(+0).
\end{equation}
It appears that this derivative has different limits for $\Delta t\to +0$ and $\Delta t\to
-0$, so the slope break at $\Delta t=0$ cannot be avoided in general.

An exception occures if $w(+0)=0$, that is if $w(t)$ decreases to zero smoothly for $\Delta
t\to +0$. This is possible if the memory function incorporates multiple minor physical
effects. In such a way, $w(t)$ represents the convolution of multiple elementary
contributions. If all of them have more or less the same magnitude and timescale, the
resulting $w(t)$ would have the desired property (smoothly vanishing at zero), and the
resulting $\kappa(\Delta t)$ would then have a high degree of smoothness. However, if just
one or two physical effects dominate over the others then $\kappa(\Delta t)$ would become
excessively peaky at zero. Then a non-smooth model might provide a better approximation.

In other words, considering only smooth GP covariances might be an unnecessary and
unjustified restriction, even if it looks reasonable on the first view.

\subsection{Deriving GP primitives from the causality principle}
\label{sec_prim}
Assume a memory function that represents a decaying sinusoid:
\begin{equation}
w(t) = {\text e}^{-\beta t} \cos\omega t, \qquad \beta>0, \quad t\geq 0.
\label{memds}
\end{equation}
Physically, $\omega$ may refer to the stellar rotation (the sinusoid is a rough
approximation of the surface rotation effect in $w$), and the exponential factor $\beta$
describes the effect of temporal decay in the surface pattern of spots/flares/etc.

Based on~(\ref{memds}), and for the stationary case $D(t)\equiv 1$, we obtain a GP with the
following characteristics:
\begin{align}
\kappa(\Delta t) &= \frac{1}{4\beta} \sqrt{\frac{\omega^2+4\beta^2}{\beta^2+\omega^2}} e^{-\beta |\Delta t|} \cos (\omega |\Delta t|+\phi), \nonumber\\
d(t) &= \frac{2\beta^2 + \omega^2}{4\beta (\beta^2+\omega^2)}, \nonumber\\
\rho(\Delta t) &= e^{-\beta |\Delta t|}\, \frac{\cos (\omega |\Delta t|+\phi)}{\cos\phi}, \nonumber\\
&\phi = \arctan\frac{\omega\beta}{2\beta^2+\omega^2}, \quad \Delta t=|t_2-t_1|.
\label{varcovexpds}
\end{align}

First of all, this general model allows to reproduce many of the special heuristic models
considered in the previous section. For $\omega=0$ this GP turns into the exponentially
correlated red noise~(\ref{covR}) with $\rho=\exp(-\beta|\Delta t|)$.

The values $\beta<0$ are forbidden, but for $\beta\to +0$ we obtain $\rho=\cos\omega \Delta
t$. This corresponds to the harmonic GP~(\ref{covH}).

A very interesting result is that in the general case, when neither $\omega$ nor $\beta$
vanish, we obtained something different from the decaying sinusoid~(\ref{covDS}). The
general expression~(\ref{varcovexpds}) remains similar in shape, but involves a phase shift
$\phi$ that depends on $\beta$ and $\omega$. This phase shift appears rather intriguing,
and offers us a theoretic possibility to observationally detect it, thus to verify
experimentally how adequate is our understanding of the ``causality restriction'' and of
the associated theory. However, possible values of $\phi$ are limited by about $20^\circ$
(attained for $\omega/\beta=\sqrt 2$), and the effect of the phase shift in $\kappa$ is
likely model-dependent. So detecting $\phi$ from the observed Doppler or photometric noise
is definitely a challenge.

Now let us consider a sinusoidal variation of the intensity $D(t)$, caused e.g. by the
stellar activity cycle:
\begin{equation}
D(t) = D_0 + D_m \cos(\Omega t + \lambda).
\label{sinmod}
\end{equation}
Assuming this $D(t)$ and the exponential $w(t)=\exp(-\beta t)$, we can obtain the following:
\begin{align}
\kappa(t_1,t_2) &= \frac{1}{2} e^{-\beta |t_2-t_1|} \left[ D_0 + D_m' \cos(\Omega \min(t_1,t_2) + \lambda') \right], \nonumber\\
d(t) &= \frac{1}{2} \left[D_0 + D_m' \cos(\Omega t + \lambda') \right], \nonumber\\
D_m' &= \frac{D_m}{\sqrt{1+\frac{\Omega^2}{4\beta^2}}},\quad \lambda' = \lambda + \arctan\frac{\Omega}{2\beta}.
\label{varcovexpsin}
\end{align}

This is somewhat different from~(\ref{covNS}). For example, the first formula contains
$(t_1+t_2)/2$ instead of $\min(t_1,t_2)$ in the second one. Again, in~(\ref{varcovexpsin})
a phase shift appears between the observed variation and the underlying activity. However,
in the case of small $\Omega/\beta$, as expected for stellar activity cycles, the
difference between these models becomes negligible.

\bibliographystyle{model2-names}
\bibliography{PlP3}

\begin{thebibliography}{40}
\expandafter\ifx\csname natexlab\endcsname\relax\def\natexlab#1{#1}\fi
\expandafter\ifx\csname url\endcsname\relax
  \def\url#1{\texttt{#1}}\fi
\expandafter\ifx\csname urlprefix\endcsname\relax\def\urlprefix{URL }\fi
\providecommand{\eprint}[2][]{\url{#2}}
\providecommand{\bibinfo}[2]{#2}
\ifx\xfnm\relax \def\xfnm[#1]{\unskip,\space#1}\fi
\bibitem[{Abubekerov and Gostev(2013)}]{AbubGost13}
\bibinfo{author}{Abubekerov, M.K.}, \bibinfo{author}{Gostev, N.Y.},
  \bibinfo{year}{2013}.
\newblock \bibinfo{title}{A universal approach to the calculation of the
  transit light curves}.
\newblock \bibinfo{journal}{\mnras} \bibinfo{volume}{432},
  \bibinfo{pages}{2216--2223}.
\bibitem[{Agol and Fabrycky(2017)}]{TTV}
\bibinfo{author}{Agol, E.}, \bibinfo{author}{Fabrycky, D.C.},
  \bibinfo{year}{2017}.
\newblock \bibinfo{title}{Transit-timing and duration variations for the
  discovery and characterization of exoplanets}, in:
  \cite{ExoplanetsHandbook}.
\bibitem[{Aigrain et~al.(2012)Aigrain, Pont and Zucker}]{Aigrain12}
\bibinfo{author}{Aigrain, S.}, \bibinfo{author}{Pont, F.},
  \bibinfo{author}{Zucker, S.}, \bibinfo{year}{2012}.
\newblock \bibinfo{title}{A simple method to estimate radial velocity
  variations due to stellar activity using photometry}.
\newblock \bibinfo{journal}{\mnras} \bibinfo{volume}{419},
  \bibinfo{pages}{3147--3158}.
\bibitem[{Anglada-Escud{\'e} et~al.(2014)Anglada-Escud{\'e}, Arriagada, Tuomi,
  Zechmeister, Jenkins, Dreizler, Gerlach, Marvin, Reiners, Jeffers, Butler,
  Vogt, Amado, Rodr{\'i}guez-L{\'o}pez, Berdi{\~n}as, Morin, Crane, Shectman,
  Thompson, D{\'i}az, Rivera, Sarmiento and Jones}]{AngladaEscude14}
\bibinfo{author}{Anglada-Escud{\'e}, G.}, \bibinfo{author}{Arriagada, P.},
  \bibinfo{author}{Tuomi, M.}, \bibinfo{author}{Zechmeister, M.},
  \bibinfo{author}{Jenkins, J.S.}, \bibinfo{author}{Dreizler, A.O.S.},
  \bibinfo{author}{Gerlach, E.}, \bibinfo{author}{Marvin, C.J.},
  \bibinfo{author}{Reiners, A.}, \bibinfo{author}{Jeffers, S.V.},
  \bibinfo{author}{Butler, R.P.}, \bibinfo{author}{Vogt, S.S.},
  \bibinfo{author}{Amado, P.J.}, \bibinfo{author}{Rodr{\'i}guez-L{\'o}pez, C.},
  \bibinfo{author}{Berdi{\~n}as, Z.M.}, \bibinfo{author}{Morin, J.},
  \bibinfo{author}{Crane, J.D.}, \bibinfo{author}{Shectman, S.A.},
  \bibinfo{author}{Thompson, I.B.}, \bibinfo{author}{D{\'i}az, M.},
  \bibinfo{author}{Rivera, E.}, \bibinfo{author}{Sarmiento, L.F.},
  \bibinfo{author}{Jones, H.R.A.}, \bibinfo{year}{2014}.
\newblock \bibinfo{title}{Two planets around kapteyn's star: a cold and a
  temperate super-{E}arth orbiting the nearest halo red-dwarf}.
\newblock \bibinfo{journal}{\mnras} \bibinfo{volume}{443},
  \bibinfo{pages}{L89--L93}.
\bibitem[{Anglada-Escud{\'e} and Tuomi(2012)}]{Anglada-Escude12}
\bibinfo{author}{Anglada-Escud{\'e}, G.}, \bibinfo{author}{Tuomi, M.},
  \bibinfo{year}{2012}.
\newblock \bibinfo{title}{A planetary system with gas giants and super-earths
  around the nearby {M} dwarf {GJ}~676{A}. {O}ptimizing data analysis
  techniques for the detection of multi-planetary systems}.
\newblock \bibinfo{journal}{\aap} \bibinfo{volume}{548}, \bibinfo{pages}{A58}.
\bibitem[{Angus et~al.(2018)Angus, Morton, Aigrain, Foreman-Mackey and
  Rajpaul}]{Angus18}
\bibinfo{author}{Angus, R.}, \bibinfo{author}{Morton, T.},
  \bibinfo{author}{Aigrain, S.}, \bibinfo{author}{Foreman-Mackey, D.},
  \bibinfo{author}{Rajpaul, V.}, \bibinfo{year}{2018}.
\newblock \bibinfo{title}{Inferring probabilistic stellar rotation periods
  using gaussian processes}.
\newblock \bibinfo{journal}{\mnras} \bibinfo{volume}{474},
  \bibinfo{pages}{2094--2108}.
\bibitem[{Baluev(2009)}]{Baluev08b}
\bibinfo{author}{Baluev, R.V.}, \bibinfo{year}{2009}.
\newblock \bibinfo{title}{Accounting for velocity jitter in planet search
  surveys}.
\newblock \bibinfo{journal}{\mnras} \bibinfo{volume}{393},
  \bibinfo{pages}{969--978}.
\bibitem[{Baluev(2011)}]{Baluev11}
\bibinfo{author}{Baluev, R.V.}, \bibinfo{year}{2011}.
\newblock \bibinfo{title}{Orbital structure of the {GJ}876 planetary system,
  based on the latest {K}eck and {HARPS} radial velocity data}.
\newblock \bibinfo{journal}{Celest. Mech. Dyn. Astron.} \bibinfo{volume}{111},
  \bibinfo{pages}{235--266}.
\bibitem[{Baluev(2013a)}]{Baluev13a}
\bibinfo{author}{Baluev, R.V.}, \bibinfo{year}{2013}a.
\newblock \bibinfo{title}{The impact of red noise in radial velocity planet
  searches: only three planets orbiting {GJ}581?}
\newblock \bibinfo{journal}{\mnras} \bibinfo{volume}{429},
  \bibinfo{pages}{2052--2068}.
\bibitem[{Baluev(2013b)}]{Baluev13c}
\bibinfo{author}{Baluev, R.V.}, \bibinfo{year}{2013}b.
\newblock \bibinfo{title}{{P}lanet{P}ack: a radial-velocity time-series
  analysis tool facilitating exoplanets detection, characterization, and
  dynamical simulations}.
\newblock \bibinfo{journal}{Astronomy \& Computing} \bibinfo{volume}{2},
  \bibinfo{pages}{18--26}.
\bibitem[{Baluev(2015)}]{Baluev14a}
\bibinfo{author}{Baluev, R.V.}, \bibinfo{year}{2015}.
\newblock \bibinfo{title}{Enhanced models for stellar {D}oppler noise reveal
  hints of a 13-year activity cycle of 55 {C}ancri}.
\newblock \bibinfo{journal}{\mnras} \bibinfo{volume}{446},
  \bibinfo{pages}{1493--1511}.
\bibitem[{Baluev and Shaidulin(2015)}]{BaluevShaidulin15}
\bibinfo{author}{Baluev, R.V.}, \bibinfo{author}{Shaidulin, V.S.},
  \bibinfo{year}{2015}.
\newblock \bibinfo{title}{Analytic models of the {Rossiter}--{McLaughlin}
  effect for arbitrary eclipser/star size ratios and arbitrary multiline
  stellar spectra}.
\newblock \bibinfo{journal}{\mnras} \bibinfo{volume}{454},
  \bibinfo{pages}{4379--4399}.
\bibitem[{Baluev et~al.(2015)Baluev, Sokov, Shaidulin, Sokova, Jones, Tuomi,
  Anglada-Escud\'e, Benni, Colazo, Schneiter, D'Angelo, Burdanov,
  Fern{\'a}ndez-Laj\'us, Ba{\c{s}}t{\"u}rk, Hentunen and Shadick}]{Baluev15a}
\bibinfo{author}{Baluev, R.V.}, \bibinfo{author}{Sokov, E.N.},
  \bibinfo{author}{Shaidulin, V.S.}, \bibinfo{author}{Sokova, I.A.},
  \bibinfo{author}{Jones, H.R.A.}, \bibinfo{author}{Tuomi, M.},
  \bibinfo{author}{Anglada-Escud\'e, G.}, \bibinfo{author}{Benni, P.},
  \bibinfo{author}{Colazo, C.A.}, \bibinfo{author}{Schneiter, M.E.},
  \bibinfo{author}{D'Angelo, C.S.V.}, \bibinfo{author}{Burdanov, A.Y.},
  \bibinfo{author}{Fern{\'a}ndez-Laj\'us, E.},
  \bibinfo{author}{Ba{\c{s}}t{\"u}rk, {\"O}.}, \bibinfo{author}{Hentunen,
  V.P.}, \bibinfo{author}{Shadick, S.}, \bibinfo{year}{2015}.
\newblock \bibinfo{title}{Benchmarking the power of amateur observatories for
  {TTV} exoplanets detection}.
\newblock \bibinfo{journal}{\mnras} \bibinfo{volume}{450},
  \bibinfo{pages}{3101--3113}.
\bibitem[{Barclay et~al.(2015)Barclay, Endl, Huber, Foreman-Mackey, Cochran,
  MacQueen, Rowe and Quintana}]{Barclay15}
\bibinfo{author}{Barclay, T.}, \bibinfo{author}{Endl, M.},
  \bibinfo{author}{Huber, D.}, \bibinfo{author}{Foreman-Mackey, D.},
  \bibinfo{author}{Cochran, W.D.}, \bibinfo{author}{MacQueen, P.J.},
  \bibinfo{author}{Rowe, J.F.}, \bibinfo{author}{Quintana, E.V.},
  \bibinfo{year}{2015}.
\newblock \bibinfo{title}{Radial velocity observations and light curve noise
  modeling confirm that kepler-91b is a giant planet orbiting a giant star}.
\newblock \bibinfo{journal}{\mnras} \bibinfo{volume}{800}, \bibinfo{pages}{46}.
\bibitem[{Barrag\'{a}n and Gandolfi(2017)}]{Barragan17a}
\bibinfo{author}{Barrag\'{a}n, O.}, \bibinfo{author}{Gandolfi, D.},
  \bibinfo{year}{2017}.
\newblock \bibinfo{title}{{\sc Exotrending}: {F}ast and easy-to-use light curve
  detrending software for exoplanets}.
\newblock \bibinfo{journal}{Astrophys. source code lib.} \eprint{ascl:
  1706.001}.
\bibitem[{Barrag\'{a}n et~al.(2017)Barrag\'{a}n, Gandolfi and
  Antoniciello}]{Barragan17b}
\bibinfo{author}{Barrag\'{a}n, O.}, \bibinfo{author}{Gandolfi, D.},
  \bibinfo{author}{Antoniciello, G.}, \bibinfo{year}{2017}.
\newblock \bibinfo{title}{{\sc pyaneti}: Multi-planet radial velocity and
  transit fitting}.
\newblock \bibinfo{journal}{Astrophys. source code lib.} \eprint{ascl:
  1706.003}.
\bibitem[{Brown et~al.(2018)Brown, Vallenari, Prusti, de~Bruijne, Babusiaux,
  Bailer-Jones et~al.}]{Gaia18}
\bibinfo{author}{Brown, A.G.A.}, \bibinfo{author}{Vallenari, A.},
  \bibinfo{author}{Prusti, T.}, \bibinfo{author}{de~Bruijne, J.H.J.},
  \bibinfo{author}{Babusiaux, C.}, \bibinfo{author}{Bailer-Jones, C.A.L.},
  et~al., \bibinfo{year}{2018}.
\newblock \bibinfo{title}{{G}aia {D}ata {R}elease 2. {S}ummary of the contents
  and survey properties}.
\newblock \bibinfo{journal}{\aa} \eprint{arXiv.org: 1804.09365}.
\bibitem[{Brown et~al.(2016)}]{Gaia17}
\bibinfo{author}{Brown, A.G.A.}, et~al., \bibinfo{year}{2016}.
\newblock \bibinfo{title}{{G}aia {D}ata {R}elease 1. {S}ummary of the
  astrometric, photometric, and survey properties}.
\newblock \bibinfo{journal}{\aap} \bibinfo{volume}{595}, \bibinfo{pages}{A2}.
\bibitem[{Deeg(2017)}]{TRVTools}
\bibinfo{author}{Deeg, H.J.}, \bibinfo{year}{2017}.
\newblock \bibinfo{title}{Tools for transit and radial velocity modelling and
  analysis}, in:  \cite{ExoplanetsHandbook}.
\bibitem[{Deeg and Belmonte(2017)}]{ExoplanetsHandbook}
\bibinfo{editor}{Deeg, H.J.}, \bibinfo{editor}{Belmonte, J.A.} (Eds.),
  \bibinfo{year}{2017}.
\newblock \bibinfo{title}{Handbook of Exoplanets}.
\newblock \bibinfo{publisher}{Springer, Cham}.
\bibitem[{Eastman et~al.(2013)Eastman, Gaudi and Agol}]{Eastman13}
\bibinfo{author}{Eastman, J.}, \bibinfo{author}{Gaudi, B.S.},
  \bibinfo{author}{Agol, E.}, \bibinfo{year}{2013}.
\newblock \bibinfo{title}{{EXOFAST}: {A} fast exoplanetary fitting suite in
  {IDL}}.
\newblock \bibinfo{journal}{\pasp} \bibinfo{volume}{125},
  \bibinfo{pages}{83--112}.
\bibitem[{Feroz and Hobson(2014)}]{FerozHobson14}
\bibinfo{author}{Feroz, F.}, \bibinfo{author}{Hobson, M.P.},
  \bibinfo{year}{2014}.
\newblock \bibinfo{title}{Bayesian analysis of radial velocity data of {GJ667C}
  with correlated noise: evidence for only two planets} \bibinfo{volume}{437},
  \bibinfo{pages}{3540--3549}.
\bibitem[{Fischer et~al.(2016)Fischer, Anglada-Escude, Arriagada, Baluev, Bean,
  Bouchy, Buchhave, Carroll, Chakraborty, Crepp, Dawson, Diddams, Dumusque,
  Eastman, Endl, Figueira, Ford, Foreman-Mackey, Fournier, F\H{u}r\'{e}sz,
  Gaudi, Gregory, Grundahl, Hatzes, H\'{e}brard, Herrero, Hogg, Howard,
  Johnson, Jorden, Jurgenson, Latham, Laughlin, Loredo, Lovis, Mahadevan,
  McCracken, Pepe, Perez, Phillips, Plavchan, Prato, Quirrenbach, Reiners,
  Robertson, Santos, Sawyer, Segransan, Sozzetti, Steinmetz, Szentgyorgyi,
  Udry, Valenti, Wang, Wittenmyer and Wright}]{Fischer16}
\bibinfo{author}{Fischer, D.A.}, \bibinfo{author}{Anglada-Escude, G.},
  \bibinfo{author}{Arriagada, P.}, \bibinfo{author}{Baluev, R.V.},
  \bibinfo{author}{Bean, J.L.}, \bibinfo{author}{Bouchy, F.},
  \bibinfo{author}{Buchhave, L.A.}, \bibinfo{author}{Carroll, T.},
  \bibinfo{author}{Chakraborty, A.}, \bibinfo{author}{Crepp, J.R.},
  \bibinfo{author}{Dawson, R.I.}, \bibinfo{author}{Diddams, S.A.},
  \bibinfo{author}{Dumusque, X.}, \bibinfo{author}{Eastman, J.D.},
  \bibinfo{author}{Endl, M.}, \bibinfo{author}{Figueira, P.},
  \bibinfo{author}{Ford, E.B.}, \bibinfo{author}{Foreman-Mackey, D.},
  \bibinfo{author}{Fournier, P.}, \bibinfo{author}{F\H{u}r\'{e}sz, G.},
  \bibinfo{author}{Gaudi, B.S.}, \bibinfo{author}{Gregory, P.C.},
  \bibinfo{author}{Grundahl, F.}, \bibinfo{author}{Hatzes, A.P.},
  \bibinfo{author}{H\'{e}brard, G.}, \bibinfo{author}{Herrero, E.},
  \bibinfo{author}{Hogg, D.W.}, \bibinfo{author}{Howard, A.W.},
  \bibinfo{author}{Johnson, J.A.}, \bibinfo{author}{Jorden, P.},
  \bibinfo{author}{Jurgenson, C.A.}, \bibinfo{author}{Latham, D.W.},
  \bibinfo{author}{Laughlin, G.}, \bibinfo{author}{Loredo, T.J.},
  \bibinfo{author}{Lovis, C.}, \bibinfo{author}{Mahadevan, S.},
  \bibinfo{author}{McCracken, T.M.}, \bibinfo{author}{Pepe, F.},
  \bibinfo{author}{Perez, M.}, \bibinfo{author}{Phillips, D.F.},
  \bibinfo{author}{Plavchan, P.P.}, \bibinfo{author}{Prato, L.},
  \bibinfo{author}{Quirrenbach, A.}, \bibinfo{author}{Reiners, A.},
  \bibinfo{author}{Robertson, P.}, \bibinfo{author}{Santos, N.C.},
  \bibinfo{author}{Sawyer, D.}, \bibinfo{author}{Segransan, D.},
  \bibinfo{author}{Sozzetti, A.}, \bibinfo{author}{Steinmetz, T.},
  \bibinfo{author}{Szentgyorgyi, A.}, \bibinfo{author}{Udry, S.},
  \bibinfo{author}{Valenti, J.A.}, \bibinfo{author}{Wang, S.X.},
  \bibinfo{author}{Wittenmyer, R.A.}, \bibinfo{author}{Wright, J.T.},
  \bibinfo{year}{2016}.
\newblock \bibinfo{title}{State of the field: Extreme precision radial
  velocities}.
\newblock \bibinfo{journal}{\pasp} \bibinfo{volume}{128},
  \bibinfo{pages}{066001}.
\bibitem[{Foreman-Mackey et~al.(2017)Foreman-Mackey, Agol, Ambikasaran and
  Angus}]{Foreman-Mackey17}
\bibinfo{author}{Foreman-Mackey, D.}, \bibinfo{author}{Agol, E.},
  \bibinfo{author}{Ambikasaran, S.}, \bibinfo{author}{Angus, R.},
  \bibinfo{year}{2017}.
\newblock \bibinfo{title}{Fast and scalable gaussian process modeling with
  applications to astronomical time series}.
\newblock \bibinfo{journal}{\aj} \bibinfo{volume}{154}, \bibinfo{pages}{220}.
\bibitem[{Fukushima(2013)}]{Fukushima13}
\bibinfo{author}{Fukushima, T.}, \bibinfo{year}{2013}.
\newblock \bibinfo{title}{Fast computation of a general complete elliptic
  integral of third kind by half and double argument transformations}.
\newblock \bibinfo{journal}{J. Comput. \& Applied Math.} \bibinfo{volume}{253},
  \bibinfo{pages}{142--157}.
\bibitem[{Gaudi and Winn(2007)}]{GaudiWinn07}
\bibinfo{author}{Gaudi, B.S.}, \bibinfo{author}{Winn, J.N.},
  \bibinfo{year}{2007}.
\newblock \bibinfo{title}{Prospects for the characterization and confirmation
  of transiting exoplanets via the {Rossiter}--{McLaughlin} effect}.
\newblock \bibinfo{journal}{\apj} \bibinfo{volume}{655},
  \bibinfo{pages}{550--563}.
\bibitem[{Hara et~al.(2017)Hara, Bou{\'e}, Laskar and Correia}]{Hara17}
\bibinfo{author}{Hara, N.C.}, \bibinfo{author}{Bou{\'e}, G.},
  \bibinfo{author}{Laskar, J.}, \bibinfo{author}{Correia, A.C.M.},
  \bibinfo{year}{2017}.
\newblock \bibinfo{title}{Radial velocity data analysis with compressed sensing
  techniques}.
\newblock \bibinfo{journal}{\mnras} \bibinfo{volume}{464},
  \bibinfo{pages}{1220--1246}.
\bibitem[{Kipping(2013)}]{Kipping13}
\bibinfo{author}{Kipping, D.M.}, \bibinfo{year}{2013}.
\newblock \bibinfo{title}{Efficient, uninformative sampling of limb darkening
  coefficients for two-parameter laws}.
\newblock \bibinfo{journal}{\mnras} \bibinfo{volume}{435},
  \bibinfo{pages}{2152--2160}.
\bibitem[{Maciejewski et~al.(2016)Maciejewski, Dimitrov, Fern\'{a}ndez, Sota,
  Nowak, Ohlert, Nikolov, Bukowiecki, Hinse, Pall'{e}, Tingley, Kjurkchieva,
  Lee and Lee}]{Maciejewski16}
\bibinfo{author}{Maciejewski, G.}, \bibinfo{author}{Dimitrov, D.},
  \bibinfo{author}{Fern\'{a}ndez, M.}, \bibinfo{author}{Sota, A.},
  \bibinfo{author}{Nowak, G.}, \bibinfo{author}{Ohlert, J.},
  \bibinfo{author}{Nikolov, G.}, \bibinfo{author}{Bukowiecki, L.},
  \bibinfo{author}{Hinse, T.C.}, \bibinfo{author}{Pall'{e}, E.},
  \bibinfo{author}{Tingley, B.}, \bibinfo{author}{Kjurkchieva, D.},
  \bibinfo{author}{Lee, J.W.}, \bibinfo{author}{Lee, C.U.},
  \bibinfo{year}{2016}.
\newblock \bibinfo{title}{Departure from the constant-period ephemeris for the
  transiting exoplanet {WASP}-12}.
\newblock \bibinfo{journal}{\aap} \bibinfo{volume}{588}, \bibinfo{pages}{L6}.
\bibitem[{Mandel and Agol(2002)}]{MandelAgol02}
\bibinfo{author}{Mandel, K.}, \bibinfo{author}{Agol, E.}, \bibinfo{year}{2002}.
\newblock \bibinfo{title}{Analytic light curves for planetary transit
  searches}.
\newblock \bibinfo{journal}{\apj} \bibinfo{volume}{580},
  \bibinfo{pages}{L171--L175}.
\bibitem[{Mayor and Queloz(1995)}]{MayorQueloz95}
\bibinfo{author}{Mayor, M.}, \bibinfo{author}{Queloz, D.},
  \bibinfo{year}{1995}.
\newblock \bibinfo{title}{A {J}upiter-mass companion to a solar-type star}.
\newblock \bibinfo{journal}{Nature} \bibinfo{volume}{378},
  \bibinfo{pages}{355--359}.
\bibitem[{Meschiari et~al.(2009)Meschiari, Wolf, Rivera, Laughlin, Vogt and
  Butler}]{Meschiari09}
\bibinfo{author}{Meschiari, S.}, \bibinfo{author}{Wolf, A.S.},
  \bibinfo{author}{Rivera, E.}, \bibinfo{author}{Laughlin, G.},
  \bibinfo{author}{Vogt, S.}, \bibinfo{author}{Butler, P.},
  \bibinfo{year}{2009}.
\newblock \bibinfo{title}{Systemic: A testbed for characterizing the detection
  of extrasolar planets. {I}. {T}he {S}ystemic {C}onsole package}.
\newblock \bibinfo{journal}{\pasp} \bibinfo{volume}{121},
  \bibinfo{pages}{1016--1027}.
\bibitem[{Nelson et~al.(2018)Nelson, Ford, Buchner, Cloutier, D\'{i}az, Faria,
  Rajpaul and Rukdee}]{Nelson18}
\bibinfo{author}{Nelson, B.E.}, \bibinfo{author}{Ford, E.B.},
  \bibinfo{author}{Buchner, J.}, \bibinfo{author}{Cloutier, R.},
  \bibinfo{author}{D\'{i}az, R.F.}, \bibinfo{author}{Faria, J.a.P.},
  \bibinfo{author}{Rajpaul, V.M.}, \bibinfo{author}{Rukdee, S.},
  \bibinfo{year}{2018}.
\newblock \bibinfo{title}{Quantifying the evidence for a planet in radial
  velocity data}.
\newblock \bibinfo{journal}{arXiv.org} \bibinfo{volume}{eprint},
  \bibinfo{pages}{1806.04683}.
\bibitem[{Olspert et~al.(2017)Olspert, Lehtinen, K{\"a}pyl{\"a}, Pelt and
  Grigorievskiy}]{Olspert18}
\bibinfo{author}{Olspert, N.}, \bibinfo{author}{Lehtinen, J.},
  \bibinfo{author}{K{\"a}pyl{\"a}, M.J.}, \bibinfo{author}{Pelt, J.},
  \bibinfo{author}{Grigorievskiy, A.}, \bibinfo{year}{2017}.
\newblock \bibinfo{title}{Estimating activity cycles with probabilistic methods
  {II}. {T}he {M}ount {W}ilson {Ca H\&K} data}.
\newblock \bibinfo{journal}{arXiv.org} \bibinfo{volume}{preprint},
  \bibinfo{pages}{1712.08240}.
\bibitem[{P{\'a}l(2010)}]{Pal10}
\bibinfo{author}{P{\'a}l, A.}, \bibinfo{year}{2010}.
\newblock \bibinfo{title}{Analysis of radial velocity variations in multiple
  planetary systems}.
\newblock \bibinfo{journal}{\mnras} \bibinfo{volume}{409},
  \bibinfo{pages}{975--984}.
\bibitem[{Patra et~al.(2017)Patra, Winn, Holman, Yu, Deming and Dai}]{Patra17}
\bibinfo{author}{Patra, K.C.}, \bibinfo{author}{Winn, J.N.},
  \bibinfo{author}{Holman, M.J.}, \bibinfo{author}{Yu, L.},
  \bibinfo{author}{Deming, D.}, \bibinfo{author}{Dai, F.},
  \bibinfo{year}{2017}.
\newblock \bibinfo{title}{The apparently decaying orbit of {WASP}-12b}.
\newblock \bibinfo{journal}{\aj} \bibinfo{volume}{154}, \bibinfo{pages}{4}.
\bibitem[{Rajpaul et~al.(2015)Rajpaul, Aigrain, Osborne, Reece and
  Roberts}]{Rajpaul15}
\bibinfo{author}{Rajpaul, V.}, \bibinfo{author}{Aigrain, S.},
  \bibinfo{author}{Osborne, M.A.}, \bibinfo{author}{Reece, S.},
  \bibinfo{author}{Roberts, S.}, \bibinfo{year}{2015}.
\newblock \bibinfo{title}{A {G}aussian process framework for modelling stellar
  activity signals in radial velocity data}.
\newblock \bibinfo{journal}{\mnras} \bibinfo{volume}{452},
  \bibinfo{pages}{2269--2291}.
\bibitem[{Rasmussen and Williams(2006)}]{GPML}
\bibinfo{author}{Rasmussen, C.E.}, \bibinfo{author}{Williams, C.K.I.},
  \bibinfo{year}{2006}.
\newblock \bibinfo{title}{Gaussian Processes for Machine Learning}.
\newblock \bibinfo{publisher}{The MIT Press}.
\bibitem[{Schneider et~al.(2011)Schneider, Dedieu, Sidaner, Savalle and
  Zolotukhin}]{Schneider11}
\bibinfo{author}{Schneider, J.}, \bibinfo{author}{Dedieu, C.},
  \bibinfo{author}{Sidaner, P.L.}, \bibinfo{author}{Savalle, R.},
  \bibinfo{author}{Zolotukhin, I.}, \bibinfo{year}{2011}.
\newblock \bibinfo{title}{Defining and cataloging exoplanets: the exoplanet.eu
  database}.
\newblock \bibinfo{journal}{\aap} \bibinfo{volume}{532}, \bibinfo{pages}{A79}.
\bibitem[{Wright and Howard(2009)}]{WrightHoward09}
\bibinfo{author}{Wright, J.T.}, \bibinfo{author}{Howard, A.W.},
  \bibinfo{year}{2009}.
\newblock \bibinfo{title}{Efficient fitting of multiplanet {K}eplerian models
  to radial velocity and astrometry data}.
\newblock \bibinfo{journal}{\apjS} \bibinfo{volume}{182},
  \bibinfo{pages}{205--215}.

\end{thebibliography}







\end{document}